\documentclass[12pt]{article}
\usepackage{epsfig}
\usepackage{amsmath}
\begin{document}
\begin{center}
\LARGE
\textbf{Causal Loops and Collapse in the Transactional
Interpretation of Quantum Mechanics}\\[1cm]
\large
\textbf{Louis Marchildon}\\[0.5cm]
\normalsize
D\'{e}partement de physique,
Universit\'{e} du Qu\'{e}bec,\\
Trois-Rivi\`{e}res, Qc.\ Canada G9A 5H7\\
email: marchild$\hspace{0.3em}a\hspace{-0.8em}
\bigcirc$uqtr.ca\\
\end{center}
\medskip
\begin{abstract}
Cramer's transactional interpretation of quantum
mechanics is reviewed, and a number of issues
related to advanced interactions and state vector
collapse are analyzed.  Where some have suggested that
Cramer's predictions may not be correct or
definite, I argue that they are, but I point
out that the classical-quantum distinction problem
in the Copenhagen interpretation has its
parallel in the transactional interpretation.
\end{abstract}
\smallskip
\small
\begin{center}
\textbf{R\'{e}sum\'{e}}
\end{center}
\begin{quote}
\makebox[5mm]{}
L'interpr\'{e}tation transactionnelle de la
m\'{e}canique quantique, propos\'{e}e par
J.~G.~Cramer, est sommairement revue, et quelques
questions li\'{e}es aux interactions avanc\'{e}es
et \`{a} l'effondrement du vecteur d'\'{e}tat
sont analys\'{e}es.  Certains ont sugg\'{e}r\'{e}
que les pr\'{e}dictions de l'interpr\'{e}tation
de Cramer ne sont pas correctes ou ne sont pas
bien d\'{e}finies.  Je cherche \`{a} montrer qu'au
contraire elles le sont, mais je signale que le
probl\`{e}me de la distinction du classique et du
quantique, inh\'{e}rent \`{a} l'interpr\'{e}tation
de Copenhague, a son parall\`{e}le dans
l'interpr\'{e}tation transactionnelle.
\end{quote}
\section{Introduction}
State vector collapse is one of the great foundational
problems of quantum mechanics.  It was postulated by
von Neumann in his theory of measurement~\cite{neumann},
as a process wherein the Schr\"{o}dinger equation
ceases to be valid.  Although von Neumann kept his
discussion to a general framework, specific models
of collapse were proposed later on, the best known
being the spontaneous localization
theories~\cite{ghirardi}.

In contradistinction with von Neumann, most approaches to
the quantum measurement problem have tried to retain
the universal validity of the Schr\"{o}dinger equation
(or of a suitable relativistic generalization).
Decoherence theory, in this context, goes a long way
towards identifying the physical variables relevant
to an effective collapsed state~\cite{schlosshauer}.
Yet many believe that decoherence by itself does not
solve the measurement problem~\cite{adler}.
Nor~\cite{marchildon} does the \emph{epistemic view}
of quantum states~\cite{peierls,peres}, according to
which the state vector represents knowledge, or
information, rather than the state of a physical system.
One way to obtain the effect of collapse without
modifying the Schr\"{o}dinger equation consists in
introducing additional structure to the Hilbert space
formalism.  Bohmian mechanics~\cite{bohm} and modal
interpretations~\cite{vermaas} implement such a program.

Cramer's transactional
interpretation~\cite{cramer1,cramer2,cramer3,cramer4,cramer5},
which has received comparatively little attention, is another
attempt to understand the effect of collapse while
keeping as close as possible to the
framework of the Schr\"{o}dinger (or some appropriate
relativistic) equation.  In Cramer's approach, a quantum
measurement or, more generally, a quantum process involving
the transfer of conserved quantities
should not be viewed as involving the
collapse of the state vector at a specific time, but rather
as a transaction involving many potential outcomes and,
most importantly, the exchange of both retarded and
advanced waves.  Cramer argues that such an outlook helps
in making sense of many otherwise paradoxical features
of quantum mechanics.

In this paper I will first give an overview of the
transactional interpretation, and then proceed to analyze
a number of issues raised in its wake.  Any theory that
introduces advanced as well as retarded interactions
raises the specter of causal loops.  In Cramer's approach
advanced waves do not allow
an observer to change its own past,
but some investigators have argued that they entail
indefinite or incorrect predictions.  I will review these
arguments, and propose a way to understand the
transactional interpretation that saves its full
predictive power.  As a by-product of the analysis, some
consequences of the atemporal view of a transaction will be
brought to light.  In the end I will argue that the
problem of the classical-quantum distinction in the
Copenhagen interpretation corresponds, in Cramer's
interpretation, to the one of making precise the
notion of a transaction.
%
\section{The transactional interpretation}
Cramer's transactional interpretation of quantum
mechanics is inspired by the electromagnetic theory
of Wheeler and Feynman~\cite{wheeler1,wheeler2}.
In this time-symmetric theory, advanced solutions
of Maxwell's equations are just as important as
retarded ones.  The electromagnetic field acting
on a charged particle comes from the other charged
particles only, and is equal to half the sum of the
retarded and advanced Li\'{e}nard-Wiechert solutions
of Maxwell's equations.  Wheeler and Feynman succeeded in
recovering the usual electromagnetic results,
including the expression for radiation reaction,
by postulating in addition that the universe is a perfect
absorber of all the electromagnetic radiation
coming from inside it.  In the Wheeler-Feynman approach,
an accelerated charge does not radiate if there are
no absorbers, and the experimentally confirmed radiation
formula is recovered if there is complete absorption.
Cramer, in the spirit of Wheeler and Feynman,
attributes physical reality both to a given
solution of the Schr\"{o}dinger equation
(propagating forward in time) and to its
complex conjugate (propagating backward in time).

An example of a quantum-mechanical interaction is the
emission of a microscopic particle (say an electron)
at some time $t_0$, followed by its absorption
at a later time $t_1$.  Originating at $t_0$,
the usual solution $\psi$ of the Schr\"{o}dinger
equation propagates through $t > t_0$, and is
called by Cramer an offer wave.  Also associated with
$t_0$ is the complex conjugate $\psi^*$, which
propagates through $t < t_0$.  The solution $\psi$
reaches all potential detectors, its amplitude
$\psi(\mathbf{r}_i, t_i)$ at detector $i$ being given
by the Schr\"{o}dinger equation.  Each detector
in turn emits both retarded and advanced (or confirmation)
waves.  Cramer argues that the confirmation wave coming
from detector $i$ reaches the source with an amplitude
proportional to
\begin{equation}
|\psi(\mathbf{r}_i, t_i)|^2
= \psi(\mathbf{r}_i, t_i) \psi^*(\mathbf{r}_i, t_i) ,
\label{conf}
\end{equation}
where the first factor on the right-hand side
coincides with the amplitude of the stimulating offer
wave at $i$, and the second one comes from the fact that
the confirmation wave develops as the time reverse of
the offer wave.
Advanced waves reaching the source in turn
stimulate a new cycle of offer and confirmation waves,
and so on.  For $t < t_0$, under appropriate boundary
conditions, the waves coming from all the
potential absorbers cancel the advanced wave coming
from the emitter.  Eventually, a transaction is
established between the emitter and one of the
detectors, and the process is complete.  If the
probability that the transaction be established with
detector $i$ is taken to be proportional to the
amplitude~(\ref{conf}) of the confirmation wave
coming from that detector, Born's quantum-mechanical
probabilistic rule follows.  (See Sec.~6
for a more complete analysis of this assertion.)

To illustrate how a transaction works, let us
consider one of the examples analyzed by Cramer,
a variant of the negative-result experiment of
Renninger~\cite{renninger} (Fig.~1).  A source S sits at the
center of a spherical shell $E_2$ of radius $R_2$.
On the interior of the shell are perfect absorbers
that will detect any particle leaving S and reaching
$E_2$.  Inside $E_2$ is a portion $E_1$ of a
concentric shell of radius $R_1$, whose interior
is also lined with perfect detectors.  The solid angle
subtended at S by $E_1$ is equal to $\Omega_1$, whereas
the solid angle subtended by the portion of $E_2$
visible from S is equal to $\Omega_2 = 4 \pi - \Omega_1$.

\begin{figure}[htb]
\begin{center}
\epsfbox{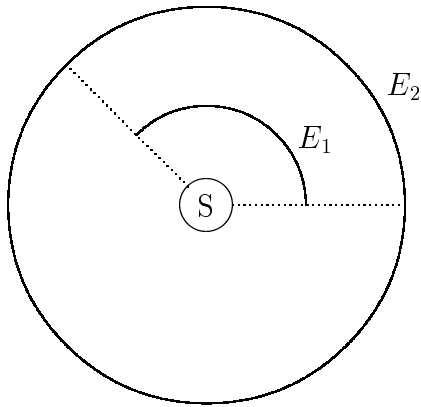}
\end{center}
\caption{A negative-result experiment.}
\end{figure}

The system is set up so that at time $t_0$, the source
emits exactly one particle with speed approximately
equal to $v$ and with a wave function independent of
angles.  For $t_0 < t < t_1 = t_0 + R_1/v$,
the state vector can be represented as
\begin{equation}
|\psi\rangle = c_1 |E_1\rangle + c_2 |E_2\rangle , 
\label{psi}
\end{equation}
where each vector $|E_i\rangle$ is associated with
detection at the corresponding shell $E_i$, and
where $|c_i|^2 = \Omega_i/4 \pi$.

Now suppose that at some time $t > t_1$, no
detector at $E_1$ has fired.  In most interpretations
of quantum mechanics, the particle's state vector is
thereafter taken to be
equal to $|E_2\rangle$.  This is certainly
good enough for the purpose of predicting results
of subsequent experiments.  Thus it seems that the
null result at $t_1$ has produced the collapse
of the state vector~(\ref{psi}) into $|E_2\rangle$.
Different views of the process of collapse may have
more or less problems in explaining how such a
mechanism can be triggered by what appears to be the
absence of a physical interaction.

In Cramer's interpretation, there is no such thing
as a collapse occurring at a specific time
$t_1$ or $t_2 = t_0 + R_2/v$.  The collapse is
interpreted as the completion of the transaction.
The whole process is viewed atemporally in
four-dimensional space-time, along the full interval
between emission and absorption.  The condition that
only one particle is emitted at the source translates
into the fact that only one transaction is
established in the end, either between S and $E_1$
(with probability $\Omega_1 /4\pi$ proportional to the
amplitude of the confirmation wave received at S from
$E_1$) or between S and $E_2$ (with corresponding
probability $\Omega_2 /4\pi$).
%
\section{Maudlin's challenge}
It is well known that in many circumstances,
advanced interactions can lead to causal paradoxes.
Such problems may be compounded in stochastic
theories, where the present doesn't uniquely
determine the future.  In quantum mechanics for
instance, the state vector at $t_0$ may specify
probabilities of measurement results at $t_1$, and
various measurement outcomes may lead to different
macroscopic configurations at $t_2$.  Hence the
configuration of absorbers at $t_2$ is not determined
by the state vector at $t_0$.  Maudlin~\cite{maudlin}
has argued that the stochasticity of the sources of
advanced waves inherent in Cramer's theory renders his
approach inconsistent.

Maudlin considers the situation depicted in
Fig.~2.  A source S can emit, at a prescribed
time $t_0$, a particle of approximate speed $v$.
The particle goes either to the left or to the
right, with equal probability.  To the right of the
source is a detector A at a distance $R_1$, and
a detector B at a larger distance $R_2$.  The
experiment is set up so that if A has not detected
a particle soon after $t_1 = t_0 + R_1/v$, detector
B is quickly moved to a distance $R_2$ to the left
of the source, in time to detect the incoming
particle at $t_2 = t_0 + R_2/v$.  (Note that a practical
realization of Maudlin's experiment would require
careful monitoring of the particle's emission time
and speed, as well as velocity spread about the two
back-to-back directions.  All of this can be
implemented within the constraints of the
uncertainty principle.)

\begin{figure}[htb]
\begin{center}
\epsfbox{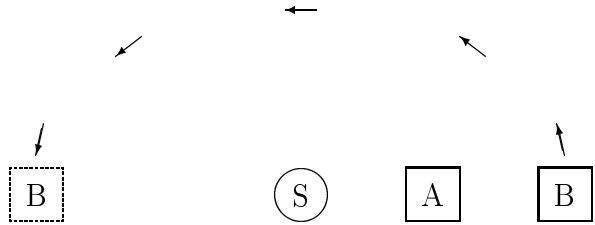}
\end{center}
\caption{Detector B quickly moves to the left
if and only if detector A does not fire.}
\end{figure}

Maudlin claims that Cramer's interpretation cannot
deal with this situation.  For the
retarded wave emitted by the source has the form
\begin{equation}
|\psi\rangle = \frac{1}{\sqrt{2}} (|L\rangle + |R\rangle) ,
\label{offer}
\end{equation}
where $|L\rangle$ and $|R\rangle$ represent states
of the particle going to the left or to the right.
Detector A, fixed on the right-hand side,
always receives the retarded wave of amplitude
$1/\sqrt{2}$, and always sends a confirmation
wave of amplitude proportional to
$(1/\sqrt{2}) (1/\sqrt{2})^* = 1/2$,
so that Cramer's theory correctly
predicts that it will absorb the particle with
probability 1/2.  But detector B receives an
offer wave and sends a confirmation wave only when
it swings to the left.  The confirmation wave again
has amplitude proportional to 1/2.
Cramer's theory therefore predicts, according to
Maudlin, that B should absorb the particle half the
times the offer wave reaches it.  But in fact it absorbs
it every time the offer wave reaches it.  Maudlin
concludes that Cramer's theory collapses.
%
\section{Berkovitz' and Kastner's answers}
Maudlin's challenge was answered in two different
ways by Berkowitz~\cite{berkovitz} and
Kastner~\cite{kastner}.  Both of them
make use of the distinction, introduced by
Butterfield~\cite{butterfield2},
between big-space and many-spaces probabilities.

Without going into all details, one can say that
the distinction is relevant in situations where
an experiment can be carried out using different
setups, or initial conditions.  Take for instance
the negative-result experiment introduced in
Sec.~2.  Suppose the experimenter decides to run
it in two different ways, either as described in
Sec.~2, or without the larger shell $E_2$.
In each run the choice of setup can be made by the
experimenter's direct intervention, or through some
automated device picking the setup either
deterministically or randomly.

In this context, big-space probabilities are defined
by considering the experimental setup $\Sigma$, as well as
the system's initial state $\Lambda$ if it is allowed
to change over runs, as random variables just like
the measurement outcome $X$.  A general probability
function $P(\Sigma , \Lambda, X)$ is therefore introduced.  
The specific probability function of outcomes for a
given setup $\Sigma_i$ and a given initial state $\Lambda_k$
is then viewed as a conditional probability, that is,
\begin{equation}
P(X|\Sigma_i, \Lambda_k) = \frac{P(\Sigma_i, \Lambda_k, X)}
{P(\Sigma_i, \Lambda_k)} ,
\end{equation}
where
\begin{equation}
P(\Sigma_i, \Lambda_k) = \sum_{\{X\}} 
P(\Sigma_i, \Lambda_k, X) . 
\end{equation}
Many-spaces probabilities, in contrast, do not
consider the setup and initial state as random
variables, and do not assign them probabilities.
Independent probability spaces are defined for
each value of the setup and initial state, the
outcome being the only random variable.  The
independent probability functions are denoted as
$P_{\Sigma_i, \Lambda_k} (X)$.

In answering Maudlin's challenge, Berkovitz first
points out that Cramer's offer waves and confirmation
waves should better be viewed as forward and
backward causal connections.  He then argues that
the confirmation waves are different in the two
cases where detector B in Fig.~2 is or is not
swung to the left.  For if B is swung, the source
receives confirmation waves from both A and B,
whereas if B is not swung, the source receives a
confirmation wave from A only.  The backward causes
being different, so should the source's initial
states be.  One can accordingly use $|\psi\rangle$ to
denote the source's initial state in cases where
B is swung to the left, and $|\psi '\rangle$ in cases
where B is not swung.

Following Butterfield, Berkovitz then argues for the
priority of many-spaces probabilities over big-space
probabilities, on the grounds that there is little
justification for assigning probabilities to states
and setups.  He then notes that in situations
where causal loops are present, one should not
expect the many-spaces probabilities (which are
properties of the setup and
state specification) to coincide
with the long-term relative frequencies.  This, he
claims, answers Maudlin's challenge, for in the
causal loop where B is swung to the left, the
long-term relative frequency of detection (equal
to~1) should not be expected to coincide with
the many-spaces probability of detection in the
initial state $|\psi\rangle$ (equal to 1/2).  There is
a price to pay for this, however.  Since one cannot
assign probabilities to $|\psi\rangle$
and $|\psi '\rangle$,
Cramer's theory can no longer be used to calculate
the unconditional relative frequencies of left
and right detection.

It is true that, in general, backward causal
connections can have the effect described by
Berkovitz.  But here the question is whether in
Cramer's theory they do or do not have that effect.
I will argue that they most likely do not.

The reason is that if we follow Berkovitz' argument,
Cramer's theory will loose predictive power in
contexts much wider than the one introduced
by Maudlin.  Consider for instance the following
two experimental setups: (i) there is only a
detector A in Fig.~2; (ii) there is a detector A
to the right and a detector B to the left in Fig.~2,
both fixed in their respective positions.  In both
cases, the source receives a confirmation wave from A,
while it receives a confirmation wave from B only in
case (ii).  If different confirmation waves mean
different causal connections in the strong sense
suggested by Berkovitz,
then Cramer's theory could not be used to predict
that A would fire with the same frequency in cases (i)
and (ii).  Indeed it could not compare any setups
where the configuration of any detector would change.
This, I submit, entails either that the causal
connections carried by the confirmation wave should
not be as strong as envisaged by
Berkovitz, or that if they are, they should be the
same irrespective of B's position.  Support for the
latter possibility will be offered in the next section.

In her response to Maudlin's challenge, Kastner
argues that Cramer's theory will keep its predictive
power intact if it uses big-space instead of
many-spaces probabilities.  This, she claims, implies
that Cramer's account of transactions occurring
in pseudotime, an account arguably not essential
to his theory, cannot be maintained.

Kastner first points out that a simple inspection
of Maudlin's setup (Fig.~2) already renders the pseudotime
account suspect.  Indeed the echoing process between
emitter and absorbers, which is supposed to determine
the absorption probabilities, presumably requires a fixed
configuration of absorbers, which is not the case
in Fig.~2.  An echo between source S
and absorber B occurs if and only if no
detection has been registered at A\@.

In Cramer's approach, when the source in
Fig.~2 emits the offer wave~(\ref{offer}), it also
emits the advanced wave
\begin{equation}
\langle\psi| = \frac{1}{\sqrt{2}} (\langle L| + \langle R|) ,
\label{adv}
\end{equation}
propagating backward in time.  If the particle
is not absorbed by A and, consequently, B swings
to the left, the confirmation waves sent by A and B
cancel the source's advanced wave at all times
$t < t_0$.  But if the particle is absorbed
by A, no confirmation wave is sent by B\@.  In this
case, Kastner argues, nothing will cancel the
$\langle L|$ part of the advanced wave emitted
by the source.  The upshot is that the total wave
configuration at times $t < t_0$ is different depending
on whether the particle is or is not absorbed by A\@.
Therefore the emission event, although it opens up
different possible futures, cannot be considered
as a branch point before which the past is fixed.

Kastner then points out that what is common to
the two cases where B does or does not swing is the
overlap between the offer and confirmation waves
between the source and A\@.  This, now taken as a
suitable branch point, allows the partitioning
of a big probability space into a subspace where
the transaction is indeed completed between the
source and A, and a subspace where it is completed
between the source and B\@.  Both are associated
with probabilities of 1/2.  Since the partition
is based on the connection between the source and A,
it does not depend in any way on an eventual echoing
between S and B\@.  The pseudotime account of the
transaction, therefore, cannot be maintained.
%
\section{A different solution}
Berkovitz and Kastner proposed two different
ways to meet Maudlin's challenge.  I will offer a
third one which, I believe, is much in the spirit
of the Wheeler-Feynman theory of advanced actions.

Let us first consider the experimental setup depicted in
Fig.~3, which is slightly different than Maudlin's.  In
fact the only difference between Fig.~2 and Fig.~3 is
that a detector C has been added to the left, farther
from the source than detector B\@.  In most
interpretations of quantum mechanics, the presence
or absence of detector C is completely immaterial,
as no particle ever reaches it.  Indeed in all cases
where a particle emitted by the source is not absorbed
by A, detector B swings to the left and absorbs it.  The
particle therefore never reaches C, and its Schr\"{o}dinger
wave function in no way depends on the presence or absence
of that additional detector.

\begin{figure}[htb]
\begin{center}
\epsfbox{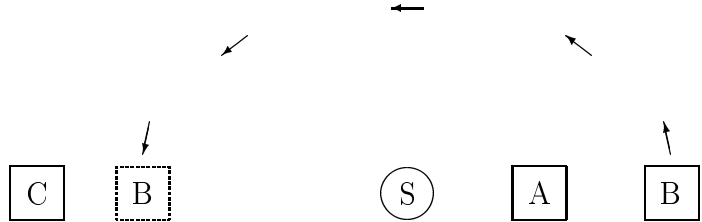}
\end{center}
\caption{Detector C will never fire.}
\end{figure}

In Cramer's interpretation, however, the presence
of C is relevant.  For in cases where the particle
is absorbed by A and B does not swing, a confirmation
wave is sent by C back to the source.  In all cases,
therefore, offer waves are sent both to the right and to
the left, and confirmation waves are received both
from the right and from the left, with amplitudes
proportional to $(1/ \sqrt{2})^2$.  As expected, the
probability of absorption on the left side is
proportional to the amplitude of the corresponding
confirmation wave.

Now how does this argument deal with the problem in
Maudlin's own situation, where no detector sits at C?
It does so by making a suitable hypothesis on the
long-distance boundary conditions.  The simplest such
hypothesis is the one made in the original approach of
Wheeler and Feynman, who postulated that the
universe is a perfect absorber of all radiation emitted
within it.  Under this assumption,
no matter that B does or does not swing,
the offer wave emitted by the source to the left
will sooner or later reach an absorber, which will send a
confirmation wave arriving at the source at $t_0$ exactly.
Just like the $\langle R|$ component in Eq.~(\ref{adv}),
the $\langle L|$ component of the advanced wave emitted
by the source will then always be cancelled by a confirmation
wave for all $t < t_0$.  The probabilities of absorption
at the left and at the right will both be proportional
to the amplitude of the corresponding confirmation
waves, that is, to 1/2.

Thus the Wheeler-Feynman hypothesis of perfect absorption
can fulfill the function of detector C in Fig.~3.  One
should point out, however, that it is not the only
assumption that will do so.  This is so much the better,
since the hypothesis has never achieved consensus.
Particles with very low interaction rates, like neutrinos,
may never find a suitable absorber.  But Cramer has
shown~\cite{cramer2} that an assumption on the
Big Bang singularity can have the same effect on these
particles as universal absorption.  The idea is
that any advanced wave reaching the $t=0$ singularity
backward from $t>0$ is assumed to be
reflected forward with a phase difference of
$180^{\circ}$.  The upshot is that the advanced wave
sent by an emitter at $t = t_0$ is cancelled for
$t < t_0$ by the reflected advanced wave, which
for $t > t_0$ reinforces the retarded wave sent by
the emitter.  To an observer who in the end
reinterprets all advanced waves moving backward in
time as retarded waves moving forward in time with
opposite momentum, reflection from the Big Bang
singularity is equivalent to perfect absorption.
 
To sum up, with either the hypothesis of perfect
absorption \emph{\`{a} la} Wheeler and Feynman, or Cramer's
assumption about the Big Bang singularity, one can
conclude that Fig.~3 correctly describes the complete
setup of Maudlin's experiment.
%
\section{Pseudotime and collapse}
In their answers to Maudlin's challenge, Berkovitz
and Kastner relied on two different interpretations of
probabilities: Berkovitz used the many-spaces approach,
whereas Kastner worked with the big space.  In the
solution proposed here, by contrast, the distinction
is not particularly critical.  The setup and initial
state can be construed as adequately specified by
the experimental protocol in place at $t_0$ (which
determines the quantum-mechanical initial wave
function).  Since the offer and confirmation waves
in place at $t_0$ do not depend on whether or not
B will swing to the left, the distinction
introduced by Berkovitz between
initial states $|\psi\rangle$ and $|\psi '\rangle$
is not relevant.  Should the experiment be run with
different setups or initial wave functions, a
many-spaces or a big-space approach could both be
put forward.

The solution proposed here also helps to clarify
the meaning of the pseudotime account of transactions.
On the assumption that the universe is a perfect
absorber (or that advanced waves are reflected at the
Big Bang), Cramer's theory correctly predicts that
in the setup of Fig.~2, absorption occurs on the left
in 50\%\ of the runs.  Yet it so happens that the
transaction is completed with the left absorber if and
only if detector B has swung to the left.  Although the
confirmation wave coming from the left originates from
a remote absorber just as often as it originates from~B,
the transaction is never completed with the remote
absorber.  It is as if the confirmation wave ``knew''
that it cannot effect a transaction in cases where
it is emitted by the remote absorber.

This can be taken as paradoxical, but I suggest it is
viewed more aptly as specifying better the significance
of a transaction.  The whole process is laid out in
four-dimensional space-time.  The future, though not
predictable, is well defined.  The offer and confirmation
waves must adapt to the requirement that the
four-dimensional process be consistent with the
boundary conditions.  In the present case, these
conditions are that B absorbs the particle if and only
if A does not absorb it.  The
forward and backward causal connections must be
strong enough to satisfy the boundary conditions,
but not so strong as to allow information transfer
or macroscopic communication.  That tension
reflects the delicate locality balance inherent
in the quantum world, as well as the transactional
interpretation's objective to avoid the notion of
collapse at a specific time.

In a general transaction, offer and confirmation
waves are exchanged between the emitter and all
possible absorbers.  It is constitutive of
the transactional interpretation that a
confirmation wave sent back from absorber~$i$ at
the space-time point $(\mathbf{r}_i, t_i)$ arrives
at the emitter with an amplitude proportional
to $|\psi(\mathbf{r}_i, t_i)|^2$.  This appears to
reproduce Born's rule with no further ado.  But
in fact the matter is more subtle.

It is well known that if $\psi(\mathbf{r}, t)$ is a
solution of the one-particle Schr\"{o}dinger equation,
the integral
\begin{equation}
P_{\text{tot}}
= \int |\psi(\mathbf{r}, t)|^2 \, d\mathbf{r} , 
\label{integral}
\end{equation}
carried over all three-dimensional
space, does not depend on time.  The
identification of $|\psi(\mathbf{r}, t)|^2$ with the
particle's position probability
density at time $t$ then implies that the
total probability is conserved, as it should.
In Cramer's approach, however, we are instructed to
associate the probability of a transaction with
$|\psi(\mathbf{r}_i, t_i)|^2$.  It turns out that in
general, an integral similar to the one on the
right-hand side of Eq.~(\ref{integral}),
but with the integrand evaluated at different times,
will not represent a conserved probability.  Does
this mean that probabilities are not conserved in
Cramer's approach?  The answer is no, for the following
reason.

Suppose that among all detectors susceptible to
send a confirmation wave back to the source, detector~1
receives the offer wave first.  Owing to the
particle-detector interaction Hamiltonian,
the particle and detector wave functions thereafter
become entangled, in a process that can be
schematized as
\begin{equation}
\psi(\mathbf{r}, t) \Phi_D^{(\text r)} \rightarrow
c_1 \psi_1(\mathbf{r}, t) \Phi_D^{(\text a)}
+ c_2 \psi_2(\mathbf{r}, t) \Phi_D^{(\text r)} .
\label{entan}
\end{equation}
Here $\Phi_D^{(\text r)}$ and $\Phi_D^{(\text a)}$ represent
ready and activated states of the detector.  The
normalized wave function $\psi_1(\mathbf{r}, t)$
evolves from the part of $\psi(\mathbf{r}, t)$ that
interacts with the detector, while the
normalized $\psi_2(\mathbf{r}, t)$ evolves
from the part that doesn't.  The coefficients
$c_1$ and $c_2$ satisfy $|c_1|^2 + |c_2|^2 = 1$.
A process like~(\ref{entan}) occurs for each further
interaction with each available detector.  In the end
probability is conserved, not because a conserved
integral similar to~(\ref{integral}) can be written
down with an integrand defined at different times,
but because the particle's reduced density matrix
always has unit trace.

This takes care of the probability question, but it
raises another issue.  As the particle outgoing from
the source interacts with various objects, its wave
function becomes entangled.  Clearly, such
entanglement is a necessary condition for a
confirmation wave to be sent back to the source,
and a transaction to be eventually completed.
But it is by no means a sufficient condition.  Any
entanglement process that is easily reversible
(i.\ e.\ one between the outgoing particle and
another microscopic object) ought not to give
rise to a confirmation wave.  For if it did lead
to a transaction, the latter would be reversible,
which is impossible.  A transaction represents
a completed quantum-mechanical process.

The question now is: What distinguishes a
reversible entanglement process, which gives
rise to no confirmation wave and no transaction,
from an irreversible one, which does give rise
to a confirmation wave and, possibly, to a
transaction.  Of course, this question has no
answer within the strict Hilbert space formalism
of quantum mechanics, where all entanglement
is in principle reversible.  The upshot is that
a transaction finds no room within the limits
of that formalism.  Just like the notion of a
classical apparatus in the Copenhagen interpretation,
or the one of wave function collapse in von
Neumann's theory of measurement, the notion of a
transaction must be added to the minimal
quantum-mechanical formalism.
In particular, the transactional
interpretation cannot be considered complete
unless the conditions for the possible
occurrence of a transaction are spelled out in
detail.

I should note that in contrasting collapse
interpretations with his own, Cramer pointed out
that collapse models ``beg the question of
borders: Where precisely is the border between
macrophysics and microphysics and the border at
which irreversibility begins?''~\cite[p.~683]{cramer3}
He seemed to imply that this problem does not show
up in the transactional interpretation.  But
elsewhere he acknowledged that ``[the transactional
interpretation's] nonlocal collapse mechanism is
strictly at the interpretational level. It cannot
supply mechanisms missing from the
formalism.''~\cite[p.~235]{cramer4}  See also
Ref.~\cite{gornitz} for additional remarks on the
Copenhagen and transactional interpretations.
%
\section{Conclusion}
Cramer's transactional interpretation uses the
notion of advanced waves in an effort to make
paradoxical quantum-mechanical processes more
intelligible.  I have argued that the
stochastic character of quantum mechanics and
the resulting unpredictability of the future
does not introduce inconsistency or loss of
predictive power in Cramer's theory.  The notion
of transaction, however, lies outside the minimal
formalism, and is in need of further
specification.
%

%
\end{document}